\documentclass[%
 reprint,
nofootinbib,
 amsmath,amssymb,
 aps,
prb,
]{revtex4-2}
\usepackage{xcolor}
\usepackage{color}
\usepackage{mathrsfs}
\usepackage{graphicx}
\usepackage{dcolumn}
\usepackage{bm}
\usepackage[normalem]{ulem}

\usepackage{hyperref}
\hypersetup{
	colorlinks = true,
	urlcolor = blue,
	linkcolor = blue,
	citecolor = green,
	filecolor = magenta,
}

\newcommand{\ud}{\mathrm{d}}
\newcommand{\pd}{\partial}
\newcommand{\sD}{{D}}
\newcommand{\zae}{{\mathcal{E}}}
\newcommand{\zam}{{\mathcal{M}}}
\newcommand{\svec}{{\mathcal{S}}}

\begin{document}

\preprint{APS/123-QED}

\title{\boldmath Gravitational Memory Effect in the Massless Vector Field }
\author{Pouneh Safarzadeh Ilkhchi$^{1}$}\email{p.safarzadeh@tabrizu.ac.ir}
\author{Amin Rezaei Akbarieh$^{1,2}$}\email{am.rezaei@tabrizu.ac.ir}
\author{Shaoqi Hou$^{3}$}\email{hou.shaoqi@whu.edu.cn}
\affiliation{$^{1}$ Faculty of Physics, University of Tabriz, Tabriz 51666-16471, Iran}
\affiliation{$^{2}$ Department of Mathematics, Faculty of Sciences, Bilkent University, 06800 Ankara, Turkey}
\affiliation{$^{3}$ School of Physics and Technology, Wuhan University, Wuhan, Hubei 430072, China}




\date{\today}

\begin{abstract}
We analyze the infrared structure and memory effects of a massless vector–tensor theory with non–minimal curvature coupling in asymptotically flat spacetimes. Using Bondi–Sachs expansions, we identify the independent radiative data and derive the effective Bondi mass aspect, whose balance law receives an additional positive definite flux from the vector sector. This leads to modified displacement, spin, and center-of-mass memory (CM) expressions, where the gravitational contributions retain their General Relativity (GR) form and the vector field enters only through well-defined flux terms. We also describe persistent vector memory effects associated with the leading angular vector mode, which are gauge-invariant but do not affect the leading tidal observables. The BMS transformations act kinematically as in GR; tensor vacua remain supertranslation degenerate, whereas the vector vacuum, defined by the vanishing vector field, is nondegenerate. All results reduce continuously to GR when the coupling is removed, isolating the precise channels through which vector–curvature interactions modify the infrared dynamics.
\end{abstract}

\maketitle


\section{Introduction}
Over the past few decades,  researchers have focused on the low-energy (infrared) features of gauge and gravity theories, due to the close link between soft theorems, asymptotic symmetries, and memory effects. \cite{Strominger:2014pwa,Strominger2014bms,Strominger:2018inf}. These developments have clarified how radiative processes imprint permanent gauge-invariant signatures on spacetime and have opened new observational and theoretical windows into the low-frequency sector of gravity and gauge fields \cite{DeLuca:2024cjl,Strominger:2018inf}.

In GR, the Gravitational Wave (GW) memory is a firm prediction that encodes both linear (ordinary) and non-linear (null) contributions arising from energy flux through null infinity \cite{Hou:2024exz,Favata:2010zu,Flanagan:2015pxa,Tachinami:2021jnf}. Recent studies have extended these ideas beyond GR, showing that modified dynamics or extra radiative Degrees of freedom(DoF) can change both the amplitude and angular pattern of memory effects, and can also lead to new types of memory not found in pure GR  \cite{Heisenberg:2023prj,Pasterski:2015tva,Du:2016hww}.
The phenomenon of GW memory has a long and rich history. Zel'dovich and Polnarev, who discovered a permanent displacement effect linked to gravitational radiation in linearized GR, are credited with the first discussions of this topic \cite{Zeldovich:1974gvh}. Later, Christodoulou showed that there is a nonlinear effect in GR that comes from the energy flux carried by the GWs themselves \cite{Christodoulou:1991cr}. Subsequent analyses by Thorne et al. clarified the observational implications and multipolar structure of the effect \cite{Thorne:1992sdb,Favata:2010zu}.

The connection between memory effects, asymptotic symmetries, and soft theorems has provided a unified framework relating classical gravitational observables to quantum infrared physics\cite{Strominger:2014pwa,Kapec:2015vwa,Pasterski:2015tva}. These developments have also led to numerous extensions of the memory framework to alternative and modified theories of gravity, including scalar-tensor and vector-tensor models \cite{Du:2016hww,Hou:2020tnd,Hou:2021oxe,Heisenberg:2023prj,Dong:2024zal,Heisenberg:2025roe}.

One of the most natural frameworks for investigating departures from Einstein gravity is the vector-tensor extension of GR, in which an additional gauge-type field couples to curvature and mediates long-range interactions \cite{Jacobson:2000xp,deFelice:2017paw}. These theories appear in a variety of contexts, including the low-energy limits of string-inspired models and effective field theories of gravity \cite{Antoniadis:1998ig, CANTATA:2021asi}. They offer an interesting framework for investigating the effects of additional DoF on gravitational radiation and its asymptotic structure \cite{deFelice:2017paw,BeltranJimenez:2016afo}.

The existence of a vector field in the infrared regime can influence the flux of energy and momentum carried to null infinity, modify the asymptotic symmetry group, and introduce new persistent features in the GW form. A convenient framework for studying how vector DoF interact with curvature and affect the infrared structure of gravity is provided by models that include a massless vector field, which can be viewed as a long-range gauge field coupled to gravity. These interactions can provide useful information on how additional gauge sectors influence asymptotic symmetries, memory effects, and soft dynamics.
By coupling geometry with vector radiation, one obtains a simple but powerful approach to explore the universality of infrared phenomena beyond the tensor regime \cite{Heisenberg:2018vsk,Aoki:2021wew,Liu:2019cxm}.
In what follows, we focus on this non-minimally coupled, massless vector model and analyze its asymptotic structure and associated memory effects in the Bondi-Sachs framework, emphasizing the modifications it induces in the infrared dynamics and the structure of asymptotic symmetries.\\

The remainder of this paper is organized as follows: in Sec.\ref{sec-vmodel}, we present the non-minimally coupled massless vector model, specify the interaction term, and derive the modified Einstein and vector field equations. Sec.\ref{sec-bsfor} introduces the Bondi-Sachs framework, including the asymptotic expansions of the metric and vector field compatible with radiative boundary conditions. In Sec.\ref{sec-asotfe}, we solve the field equations order by order near null infinity and obtain the effective Bondi mass and angular momentum aspects, identifying how the vector sector modifies the corresponding balance laws. Sec.\ref{sec-meff} is devoted to gravitational memory effects: we derive explicit expressions for the displacement, spin, and CM memories, and isolate the additional vector flux terms contributing to each. We then analyze, in Subsec.\ref{sec-vmm}, purely vector observables (electric and magnetic-type vector memories) and clarify their gauge properties and relation to the radiative vector data. Sec.\ref{sec-assym} examines the asymptotic symmetry structure and shows the BMS algebra remains kinematically identical to GR. Finally, in Sec.\ref{sec:deg-vac}, we discuss the structure of degenerate vacua and summarize our main findings and outlook in Sec.\ref{summary}.

\section{NON-MINIMALLY COUPLED VECTOR ACTION MODEL}\label{sec-vmodel}
\subsection{The Model}
We consider a class of vector–tensor theories in which the gravitational dynamics is coupled to a dynamical vector field $\zeta_\mu$. The action is written as
\begin{equation}\label{action}
    S = \int d^4x \, \sqrt{-g} \left[ f(\zeta^2) R - \frac{1}{4} F_{\mu\nu} F^{\mu\nu} \right],
\end{equation}
where $R$ is the Ricci scalar, $F_{\mu\nu} = \partial_\mu \zeta_\nu - \partial_\nu \zeta_\mu$ denotes the field strength tensor associated with the vector field, and $f(\zeta^2)$ is an arbitrary function of $\zeta^2 = \zeta_\mu \zeta^\mu$. This type of interaction encompasses a wide range of vector-tensor modifications of GR, where the vector field can interact with the curvature of spacetime. For simplicity, we will consider only this specific case \footnote{We could equivalently consider other non-minimal operators such as 
$R_{\mu\nu}\,\zeta^\mu\zeta^\nu$ or $(\nabla\!\cdot\!\zeta)^2$ (as in generalized Proca–type extensions). However, under our boundary conditions and working to the order relevant for the asymptotic expansion and memory observables, these terms do not introduce qualitatively new physics: (i) $R_{\mu\nu}\zeta^\mu\zeta^\nu$ can be traded, up to integrations by parts and field redefinitions, for a renormalization of the $R\,\zeta^2$ coupling and subleading $1/r$ corrections that do not alter the leading BMS charges and memory; (ii) the term $(\nabla\!\cdot\!\zeta)^2$ does not introduce new 
leading-order dynamics beyond that already encoded in the Maxwell kinetic term $F_{\mu\nu}F^{\mu\nu}$: it only reshuffles the longitudinal sector through field redefinitions and contributes at subleading order to the Bondi mass aspect and radiative charges. For these reasons—and to keep the minimal laboratory that still captures the vector–curvature interplay—we restrict to $f(\zeta^2)R$ with $f=\frac{1}{2\kappa}+\frac{\xi}{4}\zeta^2$.
}
\begin{equation}\label{fchoice}
    f(\zeta^2) = \frac{1}{2\kappa} + \frac{1}{4}\,\xi\, \zeta_\mu \zeta^\mu ,
\end{equation}
where $\kappa \equiv 8\pi G =M_{PL}^{-2}$ ($M_{PL}$ denotes Planck Mass), and $\xi$ is a dimensionless coupling parameter controlling the strength of the non-minimal interaction. In this case, the action reduces to \cite{Chiba:2008eh}
\begin{equation}\label{action02}
    S = \int d^4x \, \sqrt{-g} \left[ \frac{1}{2\kappa} R - \frac{1}{4} F_{\mu\nu} F^{\mu\nu} + \frac{1}{4}\, \xi R \zeta_\mu \zeta^\mu \right].
\end{equation}
The first term represents the standard Einstein-Hilbert contribution, the second is the kinetic term for the vector field, and the last term introduces a non-minimal coupling between the Ricci scalar and the norm of the vector. The non-minimal term $\frac{1}{4}\,\xi R\,\zeta_\mu\zeta^\mu$ explicitly breaks the $U(1)$ gauge symmetry 
$\zeta_\mu \!\to\! \zeta_\mu+\nabla_\mu\lambda$ because it depends on $\zeta_\mu$ rather than solely on $F_{\mu\nu}$. 
Varying the action gives a Proca-like equation, $\nabla_\nu F^{\nu\mu} \propto \xi\, R\,\zeta^{\mu}$, so curvature endows the vector with an effective, background-dependent mass $m_{\rm eff}^2 \!\propto\! \xi R$ and renders the longitudinal mode dynamical (three propagating DoF on generic backgrounds). In asymptotically flat regions where $R\!\to\!0$, this breaking becomes negligible and only the transverse radiative data on $\mathscr{I}^+$ survive; nevertheless, the coupling modifies bulk dynamics and asymptotic charges.\\ 
This non-minimally coupled vector model is a minimal deformation of GR that retains a canonical Maxwell kinetic term while allowing curvature to endow the vector with an effective mass through ($R\zeta^2$). Such a coupling captures distinctive early-universe dynamics: (i) it can support stable accelerated expansion in regimes where ($R>0$) and the longitudinal mode remains healthy \cite{Golovnev:2008cf,Heisenberg:2014rta}, (ii) it offers a controlled setting to assess how vector backgrounds preserve (or mildly break) isotropy \cite{Watanabe:2009ct}, and (iii) it cleanly separates genuinely vectorial imprints from scalar scenarios \cite{Dimopoulos:2009vu,BeltranJimenez:2016afo}. Motivated by these features, we focus on a key, observation-relevant question: how does the non-minimal vector–curvature coupling modify gravitational memory at null infinity? We derive the memory channels and charges in this model and contrast them with GR and scalar-field cases, thereby isolating signatures that could differentiate vector and scalar dynamics in the early universe.

\subsection{Field Equations}
Variation of the action \eqref{action02} with respect to the metric yields the modified Einstein equations

\begin{eqnarray}
    G_{\mu\nu}=\kappa T_{\mu\nu}\;,
\end{eqnarray}
where $T_{\mu\nu}$ denotes the effective energy-momentum tensor, which contains the vector field contributions together with the additional terms induced by the curvature coupling, and is defined as
\begin{align}\label{eq_metric}
    T_{\mu\nu}\equiv & F_{\mu\alpha} F_\nu{}^\alpha - \frac{1}{4} g_{\mu\nu} F_{\alpha\beta} F^{\alpha\beta}
    - \xi R \zeta_\mu \zeta_\nu \notag                                                                       \\
                     & + \frac{1}{2} g_{\mu\nu} \xi R \zeta_\alpha \zeta^\alpha
    - \xi R_{\mu\nu} \zeta_\alpha \zeta^\alpha \notag                                                        \\
                     & + \xi \Big( \nabla_\mu \nabla_\nu
    - g_{\mu\nu} \Box \Big) \zeta_\alpha \zeta^\alpha.
\end{align}
Note that the $R_{\mu\nu}\zeta^2$ term does not contribute at leading order in the asymptotically flat expansion near $\mathscr{I}^+$, where $R_{\mu\nu}\rightarrow 0$. This will become important when deriving the radiative data and memory. The differential operator $(\nabla_\mu\nabla_\nu - g_{\mu\nu}\Box)$ is the source of the non-trivial mixing between the curvature and the vector norm, and it will play an essential role in modifying the Bondi mass aspect and the gravitational memory effect. By variation of the action \eqref{action02} with respect to $\zeta^\nu$, we obtain the generalized vector field equation as
\begin{equation}\label{eq_vector}
    \nabla_\mu F^{\mu\nu} - \xi R \zeta^\nu = 0 ,
\end{equation}
which resembles a Proca-type equation with a curvature-dependent effective mass term \cite{Heisenberg:2014rta}.

\section{Bondi-Sachs formalism}\label{sec-bsfor}

To analyze the asymptotic dynamics and the gravitational memory effect, we adopt the Bondi-Sachs formalism. In retarded Bondi coordinates $(u,r,x^A)$, the metric is written in the form
\begin{align}\label{BondiMetric}
    ds^2 = &  e^{2\beta} \frac{V}{r} du^2 - 2 e^{2\beta} du\,dr \nonumber        \\
           & + r^2 h_{AB} \left(dx^A - U^A du \right)\left(dx^B - U^B du \right),
\end{align}

where $u$ denotes the retarded time, $r$ the radial coordinate, and $x^A$ $(A=1,2)$ angular coordinates on the unit sphere. The functions $\beta(u,r,x^A)$, $V(u,r,x^A)$, $U^A(u,r,x^A)$ and $h_{AB}(u,r,x^A)$ encode the physical DoF associated with radiation.\\

\subsection{Metric function expansion}
To ensure the flat asymptotic behavior of the metric, the metric functions must follow such expansions
\begin{eqnarray}\label{expansion_metric}
    &&\beta = \sum_{n=1}^\infty \frac{\beta_n(u,x^A)}{r^n},\nonumber\\
    && V = -r + \sum_{n=0}^\infty \frac{V_n(u,x^A)}{r^n},\nonumber\\
    &&U^A = \sum_{n=2}^\infty \frac{U^A_n(u,x^A)}{r^n},\nonumber\\
    && h_{AB} = \gamma_{AB} + \sum_{n=1}^\infty \frac{C_{AB}^{(n)}(u,x^A)}{r^n},
\end{eqnarray}
The main term in this expansion, $C_{AB}^{(1)}$, represents the propagation modes of the theory. This symmetric tensor on the sphere is decomposed into a traceless component and a trace component proportional to $\gamma_{AB}$ , where $\gamma_{AB}$ is the metric on the $2D$-unit sphere. The fall-off conditions are chosen to ensure both finiteness of the asymptotic charges and compatibility with the BMS symmetry group.

\subsection{Vector field expansion}
The asymptotic structure of the vector field must be compatible with the Bondi gauge and with the fall-off conditions of the metric. A consistent expansion is
\begin{eqnarray}\label{expansion_vector}
    \zeta_u &=& \sum_{n=1}^\infty \frac{\zeta_u^{(n)}(u,x^A)}{r^n}, \nonumber\\
    \zeta_r &=& \sum_{n=2}^\infty \frac{\zeta_r^{(n)}(u,x^A)}{r^n}, \nonumber\\
    \zeta_A &=& \sum_{n=0}^\infty \frac{\zeta_A^{(n)}(u,x^A)}{r^n}.
\end{eqnarray}
Here, the angular components $\zeta_A$ may start at order $O(r^{-1})$, while the retarded time and radial components decay more rapidly. This phrase guarantees that the stress-energy tensor derived from $\zeta_\mu$ remains finite at null infinity and that the asymptotic symmetry structure is preserved. As can be verified directly from the vector Eq. \eqref{eq_vector}, the fall-off of 
$\zeta_u$ and $\zeta_r$ follow from the $\nu=u,r$ components of $\nabla_\mu F^{\mu\nu}$, which impose $\zeta_u^{(0)}=0$ and $\zeta_r^{(0)}=\zeta_r^{(1)}=0$ for consistency with the Bondi gauge and finite energy flux at $\mathscr{I}^+$. Moreover, the energy flux component $T_{uu}$ remains finite, scaling as $T_{uu}\sim r^{-2}\,\gamma^{AB}\,\partial_u \zeta_A^{(0)}
\partial_u \zeta_B^{(0)} + O(r^{-3})$, while curvature-dependent contributions proportional to $\xi R$ vanish at null infinity where $R\rightarrow 0$. The leading angular data $\zeta_A^{(0)}$
admit the standard Hodge decomposition on the sphere, $\zeta_A^{(0)} = D_A \alpha(u,x^B) + \epsilon_A{}^{B} D_B \beta(u,x^B)$, separating the electric- and magnetic-type radiative modes. Finally, although the non-minimal coupling breaks the $U(1)$ gauge symmetry in the bulk, the asymptotic return to $R\rightarrow 0$ restores an 
effective Maxwell-like gauge freedom at $\mathscr{I}^+$, so that $\zeta_A^{(0)}$ constitutes the genuine free radiative data relevant for memory. These boundary conditions are also inspired by several works on the asymptotic analyses of Maxwell or Einstein-Maxwell theories \cite{Strominger:2018inf,Flanagan:2022pmj}.

\section{Asymptotic solution of the field equations}
\label{sec-asotfe}
In this section, we present explicit solutions of the field equations in the asymptotic region $r \to \infty$ in the Bondi-Sachs framework. In this approach, the metric functions and the vector field components are expanded in terms of inverse powers of the radial coordinate, and the resulting set of equations uniquely determines the asymptotic coefficients. Importantly, the interaction between the Bondi geometry and the vector field introduces new structures into this expansion, which we will explore in detail below. For simplicity, in the following, we will use $E_{\mu\nu}^{(n)}$ to represent the expansion coefficient of Einstein's equation at the $r^{-n}$ order. Similarly, $V_\mu^{(n)}$ is the expansion coefficient of the vector equation at the $r^{-n}$ order.

\subsection{Metric sector}
The first non-trivial information arises from the $rr$-component of the field equations. This determines the early expansion coefficients of $\beta$. The leading condition from $E_{rr}^{(3)}=0$ reads
\begin{equation}
    \beta_1 = 0,
\end{equation}
and from the equation $E_{rr}^{(4)}=0$, we obtain
\begin{equation}    
    \beta_2 = -\frac{1}{32} \, C^{(1)}_{AB} C^{AB(1)} - \frac{4}{3}\kappa \, \zeta_A^{(0)} \zeta^{(0)A}.
\end{equation}
Next, the $ur$-component equation, $E_{ur}^{(2)}=0$, implies that
\begin{equation}
    V = -r + 2 m(u,x^A)+O(r^{-1}).
\end{equation}
Here, $m(u,x^A)$ denotes an integration function known as the Bondi mass aspect, whose evolution is governed by
\begin{equation}\label{BondiMass}
    \begin{split}
        \partial_u m = & \frac{1}{4} D_A D_B N^{AB} - \frac{1}{8} N_{AB} N^{AB}                                                                          \\
                       & - \frac{1}{2}\kappa \partial_u \zeta_A^{(0)} \, \partial_u \zeta^{(0)A}-\frac{1}{2}\kappa\xi\pd_u^2(\zeta_A^{(0)}\zeta^{(0)A}),
    \end{split}
\end{equation}
as determined from the $uu$-component of the Einstein equations, i.e., $E_{uu}^{(2)}=0$.  The news tensor, denoted $N_{AB}=\partial_u C^{(1)}_{AB}$, is a traceless symmetric tensor, satisfying $N_{AB}=N_{BA}$ and $\gamma_{AB}N^{AB}=0$.
Besides the well-known terms from GR, the last contribution highlights the back reaction of the radiative vector field on the energy balance at null infinity. To make this explicit, it is convenient to define \emph{the effective Bondi mass aspect} as
\begin{equation}\label{eff-bondiMass}
    M(u,x^A)=m(u,x^A)+\kappa\xi\,\big(\zeta^{(0)}_A\pd_u\zeta^{(0)A}\big),
\end{equation}
which encodes the combined gravitational and vector contributions to the energy flux.
Then, its evolution becomes
\begin{equation}
    \label{eq-ef-bdm}
    \begin{split}
        \pd_uM= & \frac{1}{4} D_A D_B N^{AB} - \frac{1}{8} N_{AB} N^{AB}                   \\
                & - \frac{1}{2}\kappa \partial_u \zeta_A^{(0)} \, \partial_u \zeta^{(0)A},
    \end{split}
\end{equation}
where the quadratic terms can be interpreted as the energy flux densities of the tensor and the vector GWs, respectively.

The angular component $U^A$ from the $E_{rA}^{(2)}=0$ is obtained 
\begin{align}
    U_2^A & = -\frac{1}{2} D_B C^{AB(1)}, 
    \end{align}
 and from the $E_{rA}^{(3)}=0$, one can get
 \begin{align}
    U_3^A & = -\frac{2}{3} N^A + \frac{1}{3} C^{AB(1)} D^C C^{(1)}_{BC},
\end{align}
where $N^A$ is an integration function on the sphere (the angular momentum aspect). This function follows the evolution equation expressed by the $E_{uA}^{(2)}=0$,
\begin{align}
    \label{eq-evo-as}
    \partial_u N_A & =D_A M+\frac{1}{4}(D_BD_AD_CC^{BC(1)}-D_BD^BD_CC^{C(1)}_A) \notag                                                 \\
   & -\frac{1}{16}D_A(N^B_CC^{C(1)}_B)+\frac{1}{4}(N_C^BD_AC^{C(1)}_B) \notag                                                 \\
    &+\frac{1}{4}D_B(N^C_AC^{B(1)}_C-C^{C(1)}_AN_C^B)
    + \kappa\big(
    \partial_u \zeta_B^{(0)} D^B \zeta_A^{(0)} \notag                                                 \\
                   & - \partial_u \zeta^{(0)B} D_A \zeta_B^{(0)}
    - \partial_u \zeta_A^{(0)} D_B \zeta^{(0)B}\big)\notag                                            \\
                   & - \frac{3}{4}\kappa \xi D_A \, \partial_u \big( \zeta^{(0)B}\zeta^{(0)}_B \big).
\end{align}
This expression generalizes the standard Bondi flux law by introducing additional terms that depend explicitly on the leading-order vector field data, thereby modifying the angular momentum conservation at null infinity.

\subsection{Vector field sector}
\label{subs-vfs}
The set of equations for the vector field components is obtained from the variation with respect to $\zeta_\mu$. The $u$-component at order $r^{-1}$ i.e. $V_u^{(2)}=0$ gives
\begin{equation}
    \label{eq-evo-zu1}
    \pd_u\zeta_u^{(1)} = D_A\pd_u\zeta^{(0)A} - \partial_u^2 \zeta_r^{(2)}.
\end{equation}
This condition binds the subleading time component behavior of the vector to the divergence of its angular part. The  $V_u^{(3)}=0$  leads to
\begin{equation}
    \label{eq-vr-4}
    \begin{split}
         & \frac{1}{2}\zeta_r^{(2)}\pd_u^2(\zeta_A^{(0)}\zeta^A_{(0)})+\pd_u\zeta_r^{(2)}\pd_u(\zeta_A^{(0)}\zeta^A_{(0)})
        \\=&\zeta^A_{(0)}\zeta^B_{(0)}\sD_A\pd_u\zeta_B^{(0)}+\zeta^B_{(0)}\pd_u\zeta^A_{(0)}\sD_A\zeta_B^{(0)}\\
        &+\pd_u(\zeta_A^{(0)}\zeta^A_{(0)})\sD^B\zeta_B^{(0)},
    \end{split}
\end{equation}
which can be solved for $\zeta_r^{(2)}$.

For the angular components, by considering $V_A^{(2)}=0$,  we obtain
\begin{align}
    \label{eq-evo-za1}
    \partial_u \zeta^{A(1)}  =& - D^A \, \partial_u \zeta_r^{(2)}
    - \frac{1}{2}\Big( D_B D^B \zeta^{A(0)} \notag                                        \\
                            & - D_B D^A \zeta^{B(0)} - D^A D_B \zeta^{B(0)} \Big)  \notag \\
                            & + \frac{1}{2} C^{AB(1)} \partial_u \zeta^{(0)}_B
   \notag \\
    &- \frac{2}{3}\kappa \xi \, \zeta^{A(0)} \, \partial_u\big( \zeta_B^{(0)} \zeta^{(0)B} \big).
\end{align}
This equation's structure reveals pure differential constraints on the sphere, coupling with the shear \( C^{(1)}_{AB} \), and nonlinear self-interactions of the vector field mediated by curvature.
From the vector field equation $V_r^{(4)}=0$, one obtains
\begin{align}\label{eq-evo-zu2}
    \zeta_u^{(2)} & = -\frac{1}{2}( D_A \zeta^{A(1)} + D^A D_A \zeta_r^{(2)} + \partial_u \zeta_r^{(3)})\notag \\
                  & - \frac{2}{3}\,\kappa\,\xi\,\zeta_r^{(2)}\,\partial_u\big(\zeta^{(0)}_B \zeta^{B(0)}\big).
\end{align}
Here, $\zeta_u^{(2)}$ is related to $\pd_u\zeta_r^{(3)}$. 
By computing $V_u^{(4)}-\pd_uV_r^{(5)}$, one obtains a differential equation satisfied by $\zeta_r^{(3)}$. 
In addition to $\zeta_r^{(3)}$, there are $\gamma_{AB}, C_{AB}^{(1)}, \zeta_A^{(0)}$, and their derivatives in that differential equation. 
The radiative phase space of the vector sector is indeed fully captured by $\zeta_A^{(0)}(u,x)$.

Up to now, we have determined all equations satisfied by the leading order expansion coefficients $\zeta_u^{(1)}, \zeta_u^{(2)},\zeta_r^{(2)}, \zeta_r^{(3)}$, $\zeta_A^{(0)}$ ,and $\zeta_A^{(1)}$ of the vector field components. 
One can see that $\zeta_A^{(0)}$ is free, so it represents the radiative DoFs contained in the vector field $\zeta_\mu$.
This is consistent with the evolution Eq. \eqref{eq-ef-bdm} for the effective Bondi mass aspect. 
There are only two radiative DoFs provided by $\zeta_\mu$, which is consistent with the linear analysis \cite{Heisenberg:2023prj}.

To facilitate the analysis of the gravitational memory effects, we write the line element in an expanded form by substituting the asymptotic solutions for the metric functions.  The metric in retarded Bondi coordinates is
\begin{align}
\label{eq-met-exp}
    ds^2 = & -du^2 - 2\,du\,dr + r^2 \gamma_{AB}\,dx^A dx^B\notag        \\
           & + \frac{2m}{r}\,du^2
    + r\, C_{AB}^{(1)}\,dx^A dx^B
    + D_B C^{AB(1)}\,du\,dx^A \notag                                     \\
           & + \frac{4}{3r} N^A du\,dx^A
    + \frac{1}{16 r^2} C_{AB}^{(1)} C^{(1)AB}\, du\,dr\notag             \\
           & + \frac{8}{3\kappa r^2} \zeta_A^{(0)} \zeta^{(0)A}\, du\,dr
    \notag  \\
           &  + \frac{1}{4} \gamma_{AB} C^{CD(1)} C_{CD}^{(1)}\, dx^A dx^B+\cdots.
\end{align}
Although $\zeta_A^{(0)}$ shows up in the metric, the Weyl tensor has the following component,
\begin{equation}
    C_{AuBu}=-\frac{r}{2}N_{AB}+\mathcal O(r^0),
\end{equation}
which determines the geodesic deviation equation for the freely falling test particles near the null infinity.
So even though there are 4 DoF's, there are only two tensor polarizations detectable by the interferometer.
This can be easily understood because the $r$-th order of $g_{AB}$ is simply proportional to $C_{AB}^{(1)}$ as in GR.

\section{MEMORY EFFECTS}\label{sec-meff}
Having established the asymptotic structure of our non-minimally coupled vector-tensor theory, we proceed to analyze the gravitational memory effects, which constitute the primary infrared signatures of the long-range dynamics. The memory phenomenon is fundamentally linked to the asymptotic symmetries in asymptotically flat spacetimes, signifying that gravitational radiation can leave a persistent effect on the geometric data of the asymptotic region. This effect is a result of the fluxes of energy, momentum, and angular momentum transmitted to null infinity, and is properly described within the Bondi–Sachs formalism through the balance equations for the mass and angular momentum aspects. The most important manifestations of this phenomenon include the displacement memory, the spin memory, and the CM memory, each corresponding to a distinct channel of the infrared dynamics. Displacement memory originates from the persistent change in the electric-parity component of the shear data and the change in the Bondi mass aspect, whereas spin memory is associated with the magnetic-parity component and the flux of angular momentum. CM memory, conversely, appears through the dipole channel in response to the flux of linear momentum. Collectively, these three effects determine the core structure of the asymptotic transitions and reflect distinct information about the long-range radiative content of the theory. 
To clarify the memory contributions, it is useful to decompose the Bondi shear into its electric and magnetic components. Because the shear tensor is symmetric and traceless in $S^2$, it admits the decomposition\cite{Satishchandran:2019pyc}
\begin{equation}\label{decomposition}
    C^{(1)}_{AB} =
    \left(D_A D_B - \tfrac{1}{2} \gamma_{AB} D^2 \right) \Phi
    + \epsilon_{C(A} D_{B)} D^C  \Psi,
\end{equation}
where $\epsilon_{AB}$ is the antisymmetric tensor on the unit 2-sphere. Furthermore, $\Phi$ is the electric potential, and $ \Psi$ is the magnetic potential. The effect of displacement memory can be extracted from the change in $ \Phi$ using the overall conservation equation of the mass aspect.

\subsection{Displacement Memory}

In our model, the displacement memory is obtained directly from the Bondi mass loss Eq. \eqref{BondiMass}. To ensure that the time integrals appearing in the memory equations are well defined, we apply standard stationary conditions at early and late retarded times,
\begin{equation}
    \lim_{u\to \pm \infty} N_{AB}(u,x^A) = 0,
    \qquad
    \lim_{u\to \pm \infty} \partial_u \zeta^{(0)}_A(u,x^A) = 0.
\end{equation}
This assumption means that spacetime approaches non-radiative configurations as $u\to \pm\infty$, so that the Bondi mass aspect $m(u,x^A)$ tends to a finite constant value, and the integrals such as
\begin{equation}
    \int_{-\infty}^{+\infty} du \, N_{AB}N^{AB},
    \qquad
    \int_{-\infty}^{+\infty} du \, \partial_u \zeta^{(0)}_A \partial_u \zeta^{(0)A},
\end{equation}
converge.
Physically, this corresponds to the situation in which the system emits a finite burst of radiation and then settles into a new stationary state.
By integrating the Bondi mass Eq. \eqref{eq-ef-bdm} over the retarded time and rearranging terms, we obtain
\begin{align}
    D_A D_B \Delta C^{(1)AB} & = 4\,\Delta M + \frac{1}{2}\int_{-\infty}^{+\infty}du\,N_{AB}N^{AB}\notag                       \\
                             & \quad +2\kappa\int_{-\infty}^{+\infty} du \, \partial_u \zeta^{(0)}_A \partial_u \zeta^{(0)A}.
\end{align}
By substituting the shear tensor decomposition \eqref{decomposition} and considering that the magnetic-parity part is annihilated by $D_A D_B$, we obtain
\begin{align}\label{electric_part}
    D^2(D^2+2)\Delta \Phi & = 8\,\Delta M + \int_{-\infty}^{+\infty}du\,N_{AB}N^{AB}\notag                                   \\
                          & \quad + 4\kappa\int_{-\infty}^{+\infty} du \, \partial_u \zeta^{(0)}_A \partial_u \zeta^{(0)A}.
\end{align}
The above equation is obtained by applying the identity $D_A D_B\!\left(D^A D^B-\tfrac{1}{2}\gamma^{AB}D^2\right)\Phi = \tfrac{1}{2}D^2(D^2+2)\Phi$,
where $D^2 \equiv D^A D_A$ denotes the Laplacian on the unit sphere \cite{de2025lectures}. The memory potential Eq. \eqref{electric_part} can equivalently be written in the compact form
\begin{align}\label{memorysplit}
    D^2(D^2+2)\Delta \Phi & = \Delta\Phi_{\rm lin} + \Delta\Phi_{\rm nonlin}.
\end{align}
This expression explicitly shows that the displacement memory potential can be divided into two parts. The first part is the linear (ordinary) memory resulting from the change in the effective Bondi mass aspect $ M \equiv m + \kappa \xi \, \zeta_A^{(0)} \zeta^{(0)A} $ between the initial and final states. At the same time, the second part represents the nonlinear memory. This part includes both the standard gravitational-wave flux term, $\int du\,N_{AB}N^{AB}$, and a new term proportional to $\int du\,\partial_{u}\zeta^{(0)}_{A}\partial_{u}\zeta^{(0)A}$, which encodes the radiative vector field flux.
Expanding both sides of \eqref{memorysplit} in spherical harmonics $Y_{lm}$, one uses the fact that the angular operator on the left has eigenvalues
\begin{align}
    \Lambda_{l} = l(l+1)\big(l(l+1)-2\big).
\end{align}
Since $\Lambda_{l}\neq 0$ only for $l\geq 2$, the memory equation determines only the radiative multipoles\cite{Tahura:2021hbk}. For $l\geq 2$, the harmonic coefficients of the memory potential are therefore given by
\begin{align}
    \Delta\Phi_{lm}
    = \frac{\Big[\,8\Delta M
            + \int_{-\infty}^{+\infty}du\big(N_{AB}N^{AB}+4\kappa\,\partial_u \zeta^{(0)}_A \partial_u \zeta^{(0)A}\big)\Big]_{lm}}
    {\Lambda_{l}}.
\end{align}
One can see in the limit $\xi \to 0$, the effective Bondi mass reduces to the standard one, $M \to m$. In the GR limit (in the absence of a vector field) displacement memory equation becomes
\[
    \Lambda_l\, \Delta\Phi_{lm} =
    8\,\Delta m_{lm} + \Big[\int_{-\infty}^{+\infty} du \, N_{AB}N^{AB}\Big]_{lm},
\]
which is precisely the usual form of the displacement memory in GR. This confirms that our result continuously connects to the GR memory effect in the appropriate limit.
The presence of an additional vector contribution reveals important physical differences with respect to GR. Not only can the overall amplitude of the displacement memory be increased, but also the multipole structure and angular pattern of $\Delta\Phi(\theta,\phi)$ can be modified. These results confirm that in this theory, displacement memory is fully considered by computing the electric-parity potential $\Delta\Phi$.


\subsection{Spin Memory}

The spin memory effect is related to the magnetic part of the time integral of the shear tensor. To continue, we denote the magnetic potential by $\Psi(u,x^A)$ and define its initial time\cite{Tahura:2025ebb}
\begin{align}
     & \Sigma(u,x^A)\equiv\int_{-\infty}^{u} du'\,\Psi(u',x^A),\notag \\
     & \Delta\Sigma\equiv\Sigma(+\infty,x^A)-\Sigma(-\infty,x^A).
\end{align}
The quantity $\Delta\Sigma$ represents the spin memory; a nonzero $\Delta\Sigma$ indicates a permanent change in the magnetic-parity component of the shear obtained through its time integral. We start from the angular momentum flux law (\ref{eq-evo-as}),
\begin{equation}\label{NA_dot}
    \partial_u N_A = D_A M + \mathcal{G}_A[C,N] + \mathcal{V}_A[\zeta] ,
\end{equation}
where we have collected the gravitational nonlinear terms in $\mathcal{G}_A[C,N]$, exactly the same as in GR \cite{Flanagan:2015pxa}, and the vector field contributions are denoted by
\begin{align}\label{V_A_def}
    \mathcal{V}_A[\zeta]
     = &\kappa\big(\partial_u\zeta^{(0)}_B\,D^B\zeta^{(0)}_A
    -\partial_u\zeta^{(0)}_B\,D_A\zeta^{(0)B}\notag               \\
     &-\partial_u\zeta^{(0)}_A\,D_B\zeta^{(0)B}\big)
    -\tfrac{3}{4}\kappa\xi\,D_A\partial_u\big(\zeta^{(0)}_B\zeta^{(0)B}\big).
\end{align}
We now act with $\epsilon^{AB}D_B$ on \eqref{NA_dot}, in which case the $D_A M$ term vanishes, because $\epsilon^{AB}D_A D_B M=0$, so we obtain
\begin{equation}\label{curl_N_dot}
    \epsilon^{AB}D_A\partial_u N_B
    = \epsilon^{AB}D_A\mathcal{G}_B[C,N] + \epsilon^{AB}D_A\mathcal{V}_B[\zeta].
\end{equation}
Integrating the previous equation with respect to the retarded time from $u=-\infty$ to $u=+\infty$ gives the following
\begin{align}\label{curl_N_int}
    \epsilon^{AB}D_A\Delta N_B & = \int_{-\infty}^{+\infty} du\; \epsilon^{AB}D_A\mathcal{G}_B[C,N]\notag    \\
                               & \quad + \int_{-\infty}^{+\infty} du\; \epsilon^{AB}D_A\mathcal{V}_B[\zeta].
\end{align}
Using shear decomposition (\ref{decomposition}) and differential identities on the unit sphere, the left-hand side of \eqref{curl_N_int} can be related to the magnetic primitive $\Delta\Sigma$ by an elliptic operator. In particular,
\begin{align}\label{spin_id}
    \frac{1}{2}D^2(D^2+2)\,\Delta\Sigma
     =& -\,\epsilon^{AB}D_A\Delta N_B\notag                                           \\
     &  + \int_{-\infty}^{+\infty} du\; \epsilon^{AB}D_A\mathcal{F}_B[C,N,\zeta],
\end{align}
where all flux contributions are explicitly grouped within $\mathcal{F}_B$; explicitly,
\begin{align}\label{F_B}
    \mathcal{F}_B[C,N,\zeta] & =
    \mathcal{G}_B[C,N]
    \;+\;\mathcal{V}_B[\zeta],
\end{align}
with $\mathcal{V}_B[\zeta]$ given in Eq.~\eqref{V_A_def}. Eq. \eqref{spin_id} is the main equilibrium relation that determines the spin memory in terms of (i) the angular momentum aspect change $\Delta N_B$, (ii) the integrated gravitational (shear) wave flux, and (iii) the integrated vector field flux.
Finally, we decompose both sides in spherical harmonics and use the eigenvalues
\[
    \Lambda_l \equiv l(l+1)\big[l(l+1)-2\big],
    \qquad l\ge 2,
\]
and invert the angular operator to obtain the multipolar coefficients of the spin memory potential,
\begin{align}\label{DeltaSigma_lm}
    \Delta\Sigma_{lm}
     & = \frac{2}{\Lambda_l}\Big\{
    - \Big[\epsilon^{AB}D_A\Delta N_B\Big]_{lm}\notag                             \\
     & \qquad + \Big[ \int_{-\infty}^{+\infty} du\; \epsilon^{AB}D_A\mathcal{F}_B
        \Big]_{lm}
    \Big\},\qquad (l\ge 2).
\end{align}
In the GR limit, the vector field contribution to $\mathcal{V}_A[\zeta]$ vanishes, so that $\mathcal{F}_B[C,N,\zeta]\to \mathcal{G}_B[C,N]$ ($\mathcal{G}_B[C,N]$ denotes the nonlinear gravitational contribution to the angular momentum aspect.). Then Eq.~\eqref{DeltaSigma_lm} reduces to
\begin{align}
    \Delta\Sigma_{lm} & =
    \frac{2}{\Lambda_l}
    \Big\{
    - \big[\epsilon^{AB}D_A \Delta N_B\big]_{lm}\notag\\
    &+ \Big[ \int_{-\infty}^{+\infty} du \;
    \epsilon^{AB}D_A \mathcal{G}_B[C,N]\Big]_{lm}
    \Big\},
    \qquad (l\ge 2),
\end{align}
which is exactly the standard expression for the spin memory effect in GR. Eq. \eqref{DeltaSigma_lm} separates the spin memory into a linear part controlled by the change in the aspect of the angular momentum and a flux part that includes both the gravitational contribution and the vector field contributions due to $\zeta^{(0)}_A$.

\subsection{Center-of-Mass Memory}
Beyond the spin and displacement memory effects, we identify the CM memory effect, which arises from the response of the CM charge to the time-integrated fluxes at null infinity \cite{Nichols:2018qac}. 
In our model, the effective Bondi mass aspect $M$ is given in Eq.~\eqref{eff-bondiMass}, so the non-minimal coupling modifies the CM flux both through the modified mass aspect $M$ and through the vector-field contribution $\mathcal{V}_A[\zeta]$ in the angular momentum aspect law~\eqref{NA_dot}. 
To characterize the CM memory, it is convenient to introduce a scalar CM potential $\Xi$ whose change $\Delta\Xi$ encodes the part of the shear that is sourced by the \emph{divergence} of the flux \cite{Pasterski:2015tva}.\footnote{Equivalently, one may phrase CM memory in terms of the change in the CM (mass-dipole) charge in the covariant phase-space language \cite{Nichols:2018qac}.} Taking the divergence of \eqref{NA_dot}, integrating over retarded time, and using identities on $S^2$, one finds the elliptic relation
\begin{align}\label{CM_id}
    \frac{1}{2}D^2(D^2+2)\,\Delta\Xi
     =& \; D^A\Delta N_A\notag\\
    &-\; \int_{-\infty}^{+\infty} du\;(D^2M+ D^A \mathcal{F}_A[C,N,\zeta]),
\end{align}
where $\mathcal{F}_A\equiv \mathcal{G}_A+\mathcal{V}_A$ is the total (gravitational $+$ vector field) flux density introduced above. Expanding in spherical harmonics and using the same eigenvalue $\Lambda_l$ gives, for $l\ge2$,
\begin{align}\label{DeltaXi_lm}
    \Delta\Xi_{lm}
    = \frac{2}{\Lambda_l}\Big\{
    \big[D^A\Delta N_A\big]_{lm}
    -\Big[\int_{-\infty}^{+\infty}\!du\; D^A\mathcal{F}_A\Big]_{lm}
    \Big\}.
\end{align}
In particular, the $\xi$-dependence enters through $\mathcal{V}_A[\zeta]$ in Eq.~\eqref{V_A_def} and through $\Delta M$ appearing inside $\mathcal{G}_A$. In the GR limit, Eqs.~\eqref{CM_id}–\eqref{DeltaXi_lm} reduce to the standard CM-memory relations.

For completeness, we note that at the level of the metric expansion, the leading angular component of the vector field $\zeta^{(0)}_A$ also induces a correction in the $du\,dr$ sector,
\begin{equation}
    \label{eq:gur_correction}
    g_{ur}\supset \frac{8}{3}\,\kappa\,r^{-2}\,\zeta^{(0)}_A\zeta^{(0)A}\,,
\end{equation}
which reflects the backreaction on the null asymptotics in our model (coefficients consistent with the expansion used above). The CM displacement is captured by the change in the CM potential $\Delta\Xi$, projected onto the radiative $l\ge2$ multipoles. This quantity receives contributions from both the gravitational flux and the vector-field flux. From an observational viewpoint, the CM memory is expected to be small, although the additional vector-field flux term proportional to $\partial_u\zeta^{(0)}_A\partial_u\zeta^{(0)A}$ may enhance its amplitude for sufficiently large couplings \cite{Nichols:2018qac}.

\medskip
\noindent

\subsection{Vector Memory effects}
\label{sec-vmm}
If one inspects the equations for the vector field, one may discover new memory effects induced with $\zeta_A^{(0)}$.
The first equation to consider is Eq.~\eqref{eq-evo-zu1}, which looks like Eq.~\eqref{eq-ef-bdm}.
So, rearranging the terms and integrating it, one gets
\begin{equation}
    \label{eq-vmm-elec}
    D^A\Delta \zeta_A^{(2)}=\Delta\zeta_u^{(1)}+\Delta\pd_u\zeta_r^{(2)},
\end{equation}
where $\zeta_r^{(2)}$ is actually expressed in terms of $\zeta_A^{(0)}$ via Eq.~\eqref{eq-vr-4}.
Let us rewrite this equation by decomposing $\zeta_A^{(0)}$ into its electric $(\zae)$ and magnetic $(\zam)$ components,
\begin{equation}
    \label{eq-def-em-za}
    \zeta_A^{(0)}=D_A\zae+\epsilon_{AB}D^B\zam.
\end{equation}
So Eq.~\eqref{eq-vmm-elec} is equivalent to
\begin{equation}
    \label{eq-vmm-elec-2}
    D^2\Delta\zae =\Delta\zeta_u^{(1)}+\Delta\pd_u\zeta_r^{(2)}.
\end{equation}
Since this equation involves the electric component $\zae$, one may call this the electric vector memory effect.
One may also call the first term on the right-hand side the linear electric memory, and the second nonlinear.
Since $D^2$ annihilates $Y_{00}$, Eq.~\eqref{eq-vmm-elec-2} determines the change in all the spherical components $\Delta\zae_{lm}$ with $l\ge1$.

Are there more vector memories?
Let us examine Eq.~\eqref{eq-evo-za1}, which resembles Eq.~\eqref{eq-evo-as}.
So rewrite this equation by substituting Eq.~\eqref{eq-def-em-za} into the terms linear in $\zeta_A^{(0)}$, and then, one takes the divergence and the curl of the result,
\begin{gather}
    D^2D^2\zae=2D^A\pd_u\zeta_A^{(1)}+2D^2\pd_u\zeta_r^{(2)}-D^A\svec_A,\label{eq-vmm-d4e}\\
    D^2D^2\zam=2\epsilon^{AB}D_A\pd_u\zeta_B^{(1)}-\epsilon^{AB}D_A\svec_B,\label{eq-vmm-d4m}
\end{gather}
where one defines
\begin{equation}
    \label{eq-def-svec}
    \svec_A=C_{AB}^{(1)}\pd_u\zeta^{B(0)}-\frac{4}{3}\kappa\xi \zeta_A^{(0)}\pd_u(\zeta_A^{(0)}\zeta^{A(0)}).
\end{equation}
Then, integrating Eqs.~\eqref{eq-vmm-d4e} and \eqref{eq-vmm-d4m}, one obtains
\begin{gather}
    D^2D^2\int\ud u\zae=2D^A\Delta\zeta_A^{(1)}+2D^2\Delta\zeta_r^{(2)}-D^A\int\ud u\svec_A,\label{eq-vmm-d4e-i}\\
    D^2D^2\int\ud u\zam=2\epsilon^{AB}D_A\Delta\zeta_B^{(1)}-\epsilon^{AB}D_A\int\ud u\svec_B,\label{eq-vmm-d4m-i}
\end{gather}
which can be named the subleading electric and the subleading magnetic vector memory effects, respectively.
Note that there is no leading magnetic memory effect.
The existence of the electric and the magnetic memory effects also appears in other vector theories, such as Einstein-\ae{}ther theory \cite{Hou:2023pfz}.
However, the difference is that in Einstein-\ae{}ther theory, these memory effects are associated with certain symmetries.
But here, one cannot find the relevant symmetries.
So these memories may also be called the persistent variables \cite{Flanagan:2018yzh,Flanagan:2019ezo}.

Observationally, these vector memory effects cannot be detected using interferometers, as $\zeta_A^{(0)}$ does not show up in the Weyl tensor at the leading order in $1/r$. But the existence of these memory effects is unrelated to their detection.

\section{Asymptotic symmetry}\label{sec-assym}

In this section, we determine the BMS symmetries for the asymptotically flat metric \eqref{eq-met-exp} of the non-minimally coupled vector model and their kinematical action on the Bondi data and the leading vector radiative data. We then connect the symmetry action to the flux balance laws derived previously. 

As in GR, the asymptotic symmetries that preserve the Bondi gauge and fall-offs are generated by vector fields $X^\mu(\alpha,Y)$ labeled by a supertranslation function $\alpha(x^A)$ and a Conformal Killing vector (CKV) $Y^A(x^B)$ of a unit 2-sphere $(S^2,\gamma_{AB})$. 
This is because at the leading orders, the metric \eqref{eq-met-exp} takes the same form as in GR \cite{Flanagan:2015pxa}.
More explicitly, one has the following components,
\begin{align}\label{eq-def-asx}
X^u &= f,\notag \\
X^A &= Y^A - \frac{1}{r}D^A f \notag\\
&+ \frac{1}{2r^2}\,C^{(1)AB}D_B f + \mathcal{O}(r^{-3}), \notag\\
X^r &= -\frac{r}{2}\,\psi + \frac{1}{2}D^2 f
       - \frac{1}{2r}[(D_A f)D_B C^{(1)AB}\notag\\
     &  +\frac{1}{2}\,C^{(1)AB}D_A D_B f]
       + \mathcal{O}(r^{-2}),
\end{align}
where
\begin{equation}
\label{vectorfields}
f(u,x) = \alpha(x) + \frac{u}{2}\,\psi(x),\quad \psi \equiv D_A Y^A.
\end{equation}
The leading order parts are fixed by the Bondi gauge and determinant conditions and are universal, i.e., independent of the field content.

Under the infinitesimal coordinate transformation, $x^\mu\rightarrow x^\mu+X^\mu$, the metric and $\zeta_\mu$ both transform. 
Although the leading part of the metric, the first line of Eq.~\eqref{eq-met-exp}, remains the same, the higher order corrections change, still respecting the Bondi gauge and the fall-off behaviors.
Then, the kinematical actions on the shear $C^{(1)}_{AB}$, and the news tensor $N_{AB}=\partial_u C^{(1)}_{AB}$ follow from the Lie derivative $\delta_Xg_{\mu\nu}=\mathcal L_Xg_{\mu\nu}$, given by
\begin{align}
\delta_X C^{(1)}_{AB} &=  f\,N_{AB} - 2 D_A D_B f \notag\\
&+ \gamma_{AB} D^2 f
                         + \mathcal{L}_Y C^{(1)}_{AB} - \tfrac{1}{2}\,\psi\,C^{(1)}_{AB}, \label{eq:deltaC}\\
\delta_X N_{AB} &= f\,\partial_u N_{AB} + \mathcal{L}_Y N_{AB}, \label{eq:deltaN}
\end{align}
where $\mathcal L_Y$ means to take the Lie derivative on $S^2$.
Note that no explicit $\psi$-weight appears in $\delta_X N_{AB}$ because $\partial_u f=\psi/2$ cancels the conformal weight inherited from $\delta_X C^{(1)}_{AB}$ upon taking $\partial_u$. 
Note that the transformation of $C^{(1)}_{AB}$ is nonlinear, in the sense that if one starts with a trivial configuration with $C^{(1)}_{AB}=0$, one would end up with a nonvanishing shear tensor after the BMS transformation.
These are characteristic terms that indicate the nontrivial structure of the vacuum state, as illustrated in the next section.
Moreover, since $Y^A$ is a CKV, $D_AD_Bf=D_AD_B\alpha$. 
Therefore, under the Lorentz transformation, $C^{(1)}_{AB}$ transforms linearly.

Also, the variation of the Bondi mass aspect $m$ is
\begin{align}
    \delta_X m
&= f\,\partial_u m + Y^A D_A m + \tfrac{3}{2}\,\psi\, m\notag\\
&+ \tfrac{1}{4}\,N^{AB} D_A D_B f
 + \tfrac{1}{2}\,(D_A f) D_B N^{AB}\notag\\
 &+ \tfrac{1}{8}\, C^{(1)AB} D_A D_B \psi . \label{eq:deltam}
\end{align}
Thus, as in GR, the representation of BMS on $(C^{(1)}_{AB},N_{AB},m)$ is purely kinematical and universal; the vector sector affects the dynamics only through modified charges and fluxes. 
Likewise, the action of the BMS transformation on $N_A$ can also be obtained, and it is expected to depend on $\zeta_\mu$ as in Brans-Dicke theory \cite{Hou:2020tnd} and Chern-Simons theory \cite{Hou:2021oxe}.
However, the expression is too complicated and not used in the current work, so it is not given here.

The vector field $\zeta_\mu$ transforms similarly according to $\delta_X \zeta_\mu=\mathcal{L}_X\zeta_\mu$. 
With the given expressions \eqref{eq-def-asx} for $X^\mu$, there is no a prior guarantee that the asymptotic behavior of $\zeta_\mu$ remains intact. 
Luckily, as one can check, the transformed $\zeta_\mu$ still satisfies the same fall-off behavior as the original one.
This allows to determine the transformation rules of the expansion coefficients of $\zeta_\mu$.
It is easy to work out the following result for $\zeta_A^{(0)}$,
\begin{equation}
\delta_X \zeta^{(0)}_A = f\,\partial_u \zeta^{(0)}_A + \mathcal{L}_Y \zeta^{(0)}_A. \label{eq:deltazeta0}
\end{equation}
Due to the absence of $\psi$ in Eq.~\eqref{eq:deltazeta0}, $\zeta_A^{(0)}$ has a scaling weight 0.
Unlike $C^{(1)}_{AB}$, $\zeta_A^{(0)}$ transforms linearly under the infinitesimal BMS transformation. 
Therefore, the vacuum in the vector sector would be trivial.

One could also work out the transformation laws of other expansion coefficients. 
For example, one finds out that
\begin{align}
    &\delta_X\zeta_u^{(1)}=f\pd_u\zeta_u^{(1)}+\mathcal L_Y\zeta_u^{(1)}+\psi\zeta_u^{(1)}-\frac{1}{2}\zeta_A^{(0)}D^A\psi,\\
    &\delta_X\zeta_r^{(2)}=f\pd_u\zeta_r^{(2)}+\mathcal L_Y\zeta_r^{(2)}+\frac{\psi}{2}\zeta_r^{(2)}+\zeta_A^{(0)}D^Af.
\end{align}
The last terms in these equations are independent of $\zeta_u^{(1)}$ and $\zeta_r^{(2)}$, but they are never the less proportional to $\zeta_A^{(0)}$. 
So these transformations are still linear, i.e., the BMS transformation preserves a vanishing vector field.

\bigskip
\section{Degenerate vacua and BMS actions}
\label{sec:deg-vac}

The nonvanishing of $N_{AB}$ or $\pd_u\zeta_A^{(0)}$ indicates the presence of the GW near the future null infinity. 
The spacetime is thus radiative.
So a nonradiative state is defined by
\begin{equation}
N_{AB}(u,x)=0,\qquad \partial_u\zeta^{(0)}_A(u,x)=0,
\label{eq:nrstate}
\end{equation}
so that neither tensor nor vector radiation crosses $\mathscr{I}^+$.
Among all of the possible nonradiative states, there are more special ones, coined as the vacua.
Since the radiative DoF's include $C^{(1)}_{AB}$ in the tensor sector and $\zeta_A^{(0)}$ in the vector sector, there are tensor vacua and vector vacua. 

\medskip
\noindent\textbf{Tensor vacuum.} In the tensor sector, a vacuum is not simply specified by $N_{AB}=0$. 
This is due to Eq.~\eqref{eq:deltaC}.
Although by Eq.~\eqref{eq:deltaN}, $N_{AB}=0$ is preserved by a BMS transformation, $C^{(1)}_{AB}$ changes nonlinearly under the supertranslation $\alpha$, according to 
\begin{equation}
    C^{(1)}_{AB}(x^C)\rightarrow C^{(1)}_{AB}(x^C)-2D_AD_B\alpha+\gamma_{AB}D^2\alpha.
\end{equation}
So if one starts with a trivial configuration $C^{(1)}_{AB}$, which is obviously an acceptable vacuum state, one would end up with
\begin{equation}
    C'^{(1)}_{AB}=-2D_AD_B\alpha+\gamma_{AB}D^2\alpha,
\end{equation}
whose news tensor is also vanishing.
One shall realize that $C'^{(1)}_{AB}$ can actually be obtained by any infinitesimal BMS transformation, including the Lorentz transformation.
So one would propose that a generic vacuum state is characterized by a real scalar field $C(x^A)$ such that 
\begin{equation}
C^{(1)}_{AB}\big|_{\mathrm{vac}}=-2D_A D_B C+\gamma_{AB}D^2 C.
\label{eq:vacshear}
\end{equation}
$C$ transforms according to \cite{Compere:2018ylh}
\begin{equation}
    C\quad\rightarrow\quad C+\alpha,
\end{equation}
under an infinitesimal BMS transformation. 

This choice of the vacuum state agrees with the conventional definition in GR \cite{Ashtekar:1981hw,Strominger2014bms}. 
In the usual approach, one would like to use the Bondi–adapted Newman–Penrose tetrad to define the tensor vacuum according to
\begin{equation}
N_{AB}=0,\qquad \operatorname{Im}\Psi_2\big|_{1/r}=0,
\label{eq:tensorvac}
\end{equation}
where $\Psi_2$ is the Newman–Penrose Weyl scalar and $|_{1/r}$ denotes the $1/r$ coefficient in the asymptotic expansion near $\mathscr{I}^+$.
As one can check that Eq.~\eqref{eq:vacshear} satisfies Eq.~\eqref{eq:tensorvac}.
The new definition relies on the transformation properties of the radiative DoF's $C^{(1)}_{AB}$, which can be generalized to the vector sector.

A remark is in order. 
This new way of defining a vacuum state in the tensor sector does not imply Eq.~\eqref{eq:vacshear} is the unique form of the vacuum configuration. 
Indeed, a $u$-independent $C^{(1)}_{AB}$ with also the magnetic part can never the less be regarded as a valid vacuum, as in the Weyl BMS case \cite{Freidel:2021fxf}.
However, this is not usually treated as the vacuum configuration when the standard BMS symmetry is considered. 

Another way of defining the vacuum state is to solve the equations of motion by setting all of the expansion coefficients to be $u$-independent. 
Since one is interested in the vacuum state in the tensor sector, one can switch off the vector field. 
Then, the equations of motion reduces to exactly those of GR, and one can follow Ref.~\cite{Flanagan:2015pxa} to obtain the same result~\eqref{eq:vacshear}, which is in the so-called canonical frame. 
This method of identifying the vacuum can also be generalized to the vector sector.

\medskip
\noindent\textbf{Vector vacuum.}
In the vector sector, we also define the vacuum state based on the transformation property of the radiative part $\zeta_A^{(0)}$.
According to Eq.~\eqref{eq:deltazeta0}, $\zeta_A^{(0)}$ transforms linearly under an arbitrary infinitesimal BMS transformation. 
So if one starts with a nonradiative configuration \eqref{eq:nrstate}, one ends up with a different nonradiative configuration,
\begin{equation}\label{eq-tf-nrs}
    \zeta_A^{(0)}(x^B)\rightarrow\zeta'^{(0)}_A=\zeta_A^{(0)}(x^B)+\mathcal L_Y\zeta_A^{(0)}(x^B).
\end{equation}
Thus, unlike the tensor sector, the vector vacuum is nondegenerate.
Another way to understand this is the following. 
In the quantum regime, $\zeta_A^{(0)}$ acting on the state $|0\rangle$ without any vector particle creates a new state with a zero-frequency vector particle. 
After the transformation \eqref{eq-tf-nrs}, $\zeta'^{(0)}_A|0\rangle$ is still a state with a zero-frequency vector particle.
The difference is simply in the polarization of the particle. 

The form of the vacuum configuration can be fixed. 
In the vacuum state, all expansion coefficients shall be $u$-independent. 
Then, Eq.~\eqref{eq-evo-za1} gives rise to a nontrivial constraint,
\begin{equation}
    D^2\zeta_A^{(0)}-D^BD_A\zeta_B^{(0)}-D_AD^B\zeta_B^{(0)}=0,
\end{equation}
and the remaining equations listed in the previous sections hold automatically. 
With the decomposition \eqref{eq-def-em-za}, one knows that
\begin{align}
     (D^2-1)D_A\mathcal E=0,\\
    (D^2-1)D_A\mathcal M=0.
\end{align}
Therefore, both $\zae$ and $\zam$ are linearly combinations of $l=0$ spherical Harmonics.
So in the vacuum state,
\begin{eqnarray}
    \zeta_A^{(0)}=0,
\end{eqnarray}
as in the Maxwell theory.
This configuration is invariant under the BMS transformation, which also explains the nondegeneracy of the vector vacuum.

\medskip
\noindent\textbf{Vacuum transitions and memory.}
As discussed in the above, the vacuum is degenerate only in the tensor sector.
So the vacuum transition occurs only in the tensor sector. 
There would be no vacuum transition in the vector sector. 

Transitions between tensor vacua are encoded by displacement memory: using the effective mass aspect and its balance law, the electric-parity memory potential is fixed by \eqref{electric_part}. 
Spin memory follows from the angular momentum aspect Eq. \eqref{NA_dot} together with the curl identity \eqref{spin_id}, which determines the magnetic-parity primitive $\Delta\Sigma$ up to the kernel of $D^2(D^2+2)$ (removing the $\ell=0,1$ modes). The CM memory arises from the divergence of \eqref{NA_dot} and is determined by \eqref{CM_id} for the even-parity CM potential $\Delta\Xi$. In all three channels, the only departures from GR are the vector-sector flux $\mathcal{V}_A[\zeta]$ entering \eqref{NA_dot} and the $\xi$-dependent contribution to the effective mass aspect feeding \eqref{electric_part}; the GR limit is recovered smoothly as $\xi\to0$.

Finally, the leading  vector memory effect~\eqref{eq-vmm-elec-2} looks like the tensor displacement memory, but it is not related to any vacuum transition in the vector sector. 
It is simply a transition between nonradiative configurations.

\section{Summary and Conclusion}\label{summary}
We analyzed a non–minimally coupled, massless vector–tensor theory in asymptotically flat spacetimes using the Bondi–Sachs framework. Starting from an action where the vector couples to curvature, we derived the modified Einstein and vector equations, imposed consistent large-$r$ falloffs for the metric and vector, and solved the hierarchy of constraints to identify the independent radiative data: the Bondi shear $C^{(1)}_{AB}$ and the leading angular mode $\zeta^{(0)}_A$. A central output is an \emph{effective} Bondi mass aspect that packages metric and vector contributions and whose balance law retains the GR structure while acquiring a positive-definite vector flux built from $\partial_u \zeta^{(0)}_A$. This provides a compact way to track how the vector sector feeds energy through null infinity.

Within this setup, we derived closed memory relations. The displacement (electric-parity) memory is sourced by the change of the effective mass aspect together with the sum of tensor and vector fluxes. The spin (magnetic-parity) memory follows from the curl channel of the angular momentum balance, and the CM memory from the divergence channel; in both, the vector sector enters through well-defined flux densities. We also exhibited vector-specific persistent observables tied to $\zeta^{(0)}_A$ at leading and subleading orders. These are gauge-invariant asymptotic effects that do not modify the leading $1/r$ Weyl curvature and therefore sit outside standard interferometric tidal observables.

Finally, we examined the BMS action on the asymptotic data. The kinematical transformation laws for $(C^{(1)}_{AB},N_{AB},m)$ coincide with GR, so the symmetry algebra and its representation remain unchanged, while $\zeta^{(0)}_A$ transforms as a weight-zero covector on $S^2$. Vacuum structure reflects this split: tensor vacua are supertranslation-degenerate (labeled by a scalar $C$ on the sphere), whereas with the standard nonradiative condition $\partial_u\zeta^{(0)}_A=0$ the vector vacuum is nondegenerate under conformal Killing vectors. All formulas reduce smoothly to GR when the non-minimal coupling is removed, isolating the precise infrared channels—energy balance and fluxes—through which vector–curvature interactions leave signatures.

\begin{acknowledgments}

	S. H. was supported by the National Natural Science Foundation of China under Grant No.~12205222.

\end{acknowledgments}

\bibliography{refs}

\end{document}